\documentclass[aps,pre,twocolumn,showpacs,superscriptaddress]{revtex4-1}

\usepackage{bm}%
\usepackage{float}
\usepackage{dsfont}
\usepackage[T1]{fontenc}
\usepackage[pdftex]{graphicx}
\usepackage{epsfig}
\usepackage{graphicx}
\usepackage{dcolumn}
\usepackage{bm}
\usepackage{color}
\newcommand {\be}{\begin{equation}}
\newcommand {\ee}{\end{equation}}
\newcommand {\ba}{\begin{eqnarray}}
\newcommand {\ea}{\end{eqnarray}}
\usepackage{amsmath,amssymb}  
\usepackage{amsfonts} 
\usepackage{graphicx} 
\usepackage{nicefrac}
\usepackage{hyperref}
\usepackage{epstopdf}
\usepackage{enumitem}
\usepackage[normalem]{ulem}
\usepackage{hyperref}
\begin{document}


\title{Non-Markovian entropy production fluctuation theorem driven by a time-dependent electric field} 
\author{K. S. Rodr\'iguez-Vigil}
\author{M. A. Bastarrachea-Magnani}
\author{J. I. Jim\'enez-Aquino} 
\affiliation{Departamento de F\'isica, Universidad Aut\'onoma Metropolitana-Iztapalapa, Av. Ferrocarril San Rafael Atlixco 186, C.P. 09310 CDMX, M\'{e}xico}

\begin{abstract}
Fluctuation theorems are key to understanding both fundamental and applied aspects of non-equilibrium thermodynamics of small systems. We study the non-Markovian entropy production fluctuation theorem for the diffusion process of charged particles in a gas inside a harmonic potential and under the action of a time-dependent electric field, using a generalized Langevin equation. By considering the influence of the electric field on both the tagged Brownian particle and the bath particles, an ``induced" electric force arises. Despite the additional force, we demonstrate that Kubo's second fluctuation-dissipation theorem (FDT) remains unchanged. The FDT allows us to obtain the Gaussian probability density for the position along a single stochastic trajectory, which is the key to demonstrating the validity of the detailed fluctuation theorem (DFT) for the total entropy production. We study the specific result of an Ornstein-Uhlenbeck-type friction memory kernel and an oscillating electric field, and analyze the average work and entropy production in different parameter regimes. 
\end{abstract}

\maketitle 

\section{Introduction}
\label{sec:1}

The 1980s and 1990s witnessed the emergence of key thermodynamic relationships that have revolutionized the statistical physics of non-equilibrium systems, where stochastic fluctuations play a fundamental role. We talk about the well-known fluctuation theorems (FTs), discovered in several pioneering works~\cite{Bochkov1981,Evans1993,Gallavotti1995,Jarzynski1997,Crooks1999,Kurchan1999}, which have improved our understanding of the second law of thermodynamics and the emergence of irreversibility from time-reversible microscopic equations of motion. The FTs relate probability distribution functions along forward and backward trajectories for small fluctuating systems. They have been extended to other thermodynamic quantities like work, heat, power, entropy production, among others~\cite{vanZon2004,Noh2012,Jimenez2013,Rao2018,Park2023,Semeraro2024}. The FTs continue to be interesting for describing fluctuating phenomena in novel systems, such as the quantum regime~\cite{Funo2018,Binder2018}, biophysical systems~\cite{Hayashi2018}, active matter~\cite{Chaki2018,Goswami2019,Narinder2021,Argun2016}, topological systems~\cite{Mahault2022}, the Maxwell's demon~\cite{Zeng2021},  human sensorimotor learning~\cite{Hack2023}, and even the UNO card game~\cite{Sidajaya2025}.

In the context of stochastic thermodynamics, entropy can be consistently defined along a single stochastic trajectory. This is done in Ref. \cite{Seifert2005} in which the transient integral fluctuation theorem (IFT) for the total entropy production, that is, $\langle e^{-\Delta s_{tot}}\rangle=1$, is shown to be valid, where $\Delta s_{tot}$ involves both the system's entropy (a Brownian particle) and the entropy production of the surroundings. Likewise, it was also shown that the detailed fluctuation theorem (DFT) is valid in the nonequilibrium steady state over a finite time interval, i.e., $P(\Delta s_{tot})/P(-\Delta s_{tot})=e^{\Delta s_{tot}}$, where $P(\pm\Delta s_{tot})$ is the probability for the total entropy production along a forward (backward) trajectory. Almost four years later, Seifert's proposal could be extended to demonstrate the validity of the DFT for the total entropy production in the transient case, if the initial probability distribution satisfies the equilibrium canonical one ~\cite{Saha2009}. In a similar way, the study in~\cite{Saha2009} could also be extended to demonstrate the validity of the total entropy production for a Brownian particle in crossed electric and magnetic fields~\cite{Jimenez2010}.

The studies reported in Refs.~\cite{Seifert2005,Saha2009,Jimenez2010}, were formulated for Markovian stochastic systems, characterized by a standard Langevin equation with additive Gaussian white noise. However, the study of the FT, particularly the Jarzynski and Crooks theorems, have been extended to non-Markovian baths~\cite{Mai2007}, and those governed by a non-linear generalized Langevin equation (GLE) \cite{Ohkuma2007}. Also, Jarzynski and Crooks theorems were proved to be valid for a charged Brownian particle bounded to a harmonic potential trap and under the action of crossed electric and magnetic fields~\cite{Jimenez2015}. In 2017, the transient DFT for the total entropy production was extended to non-Markovian systems in Ref.~\cite{Ghosh2017}, in which the solution strategy is based on the GLE and the Fokker-Planck equation associated with the probability density $P(x,w,t)$, being $x$ the Brownian particle position, $w$ the thermodynamic work along a stochastic trajectory, and $t$ the time. 

The purpose of this work is to prove the validity of the DFT for the total entropy production for a gas of charged particles performing Brownian motion under the action of a time-dependent electric field, which influences both the tagged Brownian particle (BP) and the bath particles. The GLE for this system has been derived before~\cite{Cui2018} in the context of the Zwanzig-Caldeira-Leggett model~\cite{Zwanzig2001}, and  due to the influence of the electric field on the bath particles, an additional induced force arises in the GLE. There, the authors suggest that Kubo's second FDT must be modified ~\cite{Cui2018}, that is, the noise correlation function is not only proportional to the friction memory kernel, but it also contains an additional contribution  proportional to the product of two electric fields, each at different times. Here, our strategic solution is simpler and is based directly on the explicit solution of the GLE, assuming that the system is bound to a harmonic potential trap. This allows us to obtain the Gaussian probability density (GPD), $P(x,t)$, a key component in proving the validity of the DFT for the total entropy production. 

We propose two ways to calculate the GPD: one is when the induced electric force ${\mathcal F}_{ein}(t)$ is added to the external electric force ${\mathcal F}_e(t)=QE(t)$ ($Q$ being the Brownian particle charge) resulting in an {\it effective electric force} $F_e(t)={\mathcal F}_e(t)+{\mathcal F}_{ein}(t)$. Next, we use Zwanzig's method to show that the colored noise correlation function satisfies Kubo's second FDT. The GPD is obtained by solving the GLE, along with the use of Kubo's second FDT. The other way is when the induced electric force ${\mathcal F}_{ein}(t)$ is now added to the internal noise $\eta(t)$, resulting in an {\it effective internal noise} $\eta_{\text{eff}}(t)={\mathcal F}_{ein}(t)+\eta(t)$. This proposal suggests a modification of Kubo's second FDT as reported in \cite{Cui2018}. However, despite this proposal, we will show that the GPD is the same as the one obtained using Kubo's theorem.  Therefore, both strategies lead to the same GPD, which explains why, even in the presence of the electric force plus the induced one in the GLE, the usual Kubo's second FDT remains valid. 

On the other hand, it is well known that in the FTs, the time-dependent protocol is identified with the external force that drives the system out of equilibrium. In the present work, this is precisely the role played by the effective force $F_e(t)$ defined above. So, the validity of total entropy DFT will be studied for two specific protocols: (i) the case of a time-dependent external field, and (ii) the dragging of the minimum of the potential trap. The theoretical results will be applied to the case of an oscillating electric field for which the induced electric force has been shown to be proportional to the electric field under certain physical considerations~\cite{Cui2018}. Therefore, the effective force is also $F_e(t)\propto E(t)$. Additionally, we include a friction memory kernel that satisfies an exponentially decaying function of the Ornstein-Uhlenbeck (OU) type, so that we explicitly calculate the average of both the work and the total entropy production across the friction vs. driving frequency parameter space.

The article is organized as follows. In Sec.~\ref{sec:2}, we use the Zwanzig-Caldeira-Leggett model to derive the GLE. By proposing the definition of an effective force, we show that Kubo's second FDT remains unchanged.  Next, we propose an alternative method based on an effective noise that allows us to obtain the same modified Kubo's second FDT reported in Ref.~\cite{Cui2018}. Section~\ref{sec:3} is devoted to explicitly solving the GLE for a system bound to a harmonic potential trap and under an arbitrary time-dependent effective force. The GPD is calculated by taking into account Kubo's second FDT and its modified version, yielding the same result in both cases. Sec.~\ref {sec:4} focuses on demonstrating the validity of total entropy production DFT for the two protocols aforementioned, and in Sec.~\ref{sec:5} we calculate the mean work and total entropy production as a function of the dynamic parameters of the system. Finally, in Sec.~\ref{sec:6} we offer our Conclusions. We also include three Appendices with explicit calculations.



\section{Particle-bath system under time-dependent force}
\label{sec:2}

We consider a Zwanzig-Caldeira-Leggett (ZCL) model under a time-dependent external field. It entails a Brownian particle (BP) with mass $m$ and electric charge $Q$, embedded in a charged fluid, which serves as a bath, constituted of a large number of independent particles acting as harmonic oscillators each with mass $m_{i}$, charge $q_{i}$, and frequency $\omega_{i}$. The Hamiltonian reads~\cite{Zwanzig2001,Cui2018,Lisy2019}
\begin{gather} \label{He} 
H={p^2\over 2m}+{\mathcal V}_{0}(x)+\sum_{i} {p^2_i\over 2m_i}\\ \nonumber
+\sum_i \left[{m_i\omega_i^2\over 2}\left(x_{i} -{c_{i}\over\omega^2_{i}} x\right)^2\right]
- Q\,E(t)\left(x+\sum_{i} \frac{q_i}{Q} x_{i} \right),          
\end{gather}
where the first line accounts for the non-interacting energy with the momenta along the $x$-axis for the tagged and bath particles being $p$ and $p_i$, respectively, and $\mathcal{V}_{0}(x)$ accounts for the potential energy associated with any other conservative force (either linear or non-linear) over the BP. The tagged BP is linearly coupled to the bath oscillators with coupling strength $c_i$. Finally, the particle-bath system is under the action of a time-dependent electric field $E(t)$ pointing along the $x$-axis of a Cartesian coordinate frame. This Hamiltonian can model a system of charged or polarizable particles, either ions in plasma, electrons and ions in liquid metal, or pairs of negatively and positively charged particles representing the electron cloud and the molecular ion of a polarized neutral molecule, as, for example, in the dielectric relaxation of molecular liquids~\cite{Cui2018}. Accordingly, Hamilton's equations are~\cite{Zwanzig2001}
\begin{eqnarray}
\label{dxdp} 
\dot x&=& p/m, \\   
\dot p&=&-{\mathcal V}_0^{\prime}(x)+ \sum_i m_i c_i \left(x_i - {c_i\over \omega_i^2} x\right) +Q E(t), \\ 
\label{dxidpi}
\dot x_{i}&=& p_{i}/m_{i},\\
 \dot p_i&=&m_{i}c_{i}\left(x-\frac{\omega_i^2}{c_{i}} x_{i} \right) +q_iE(t) .  
\end{eqnarray}
Using the Laplace transform to solve the differential equation for $\ddot x_i$, and the subsequent algebra, we arrive to the GLE~\cite{Cui2018,Lisy2019} 
\ba
m \dot v(t)&=&-{\mathcal V}_0^{\prime}(x) -\int_0^t \gamma(t-t^{\prime})  \, v(t^{\prime}) \, dt^{\prime} + {\mathcal F}_{ein}(t) \cr\cr
&+& Q E(t) +\eta(t),  \label{gle1} 
\ea
where  
\begin{equation}
{\mathcal F}_{ein}(t)=\sum_i {q_i c_i \over \omega_i} \int_0^t E(t^{\prime})  \sin[\omega_i (t-t^{\prime})] \, dt^{\prime} ,\label{F} 
 \end{equation}
is the induced electric force coming from the thermal interaction between the applied electric field and bath particles, $\gamma(t-t^{\prime})$ is the generalized friction memory kernel given by
 \begin{equation}
 \gamma(t-t^{\prime})= \sum_i  {m_ic^2_i \over\omega_i^2} \cos[\omega_i (t-t^{\prime})], \label{g}
\end{equation}
and $\eta(t)$ the internal noise which reads as  
\ba
 \eta(t)&=&\sum_i \bigg\{  m_i c_i \left(x_i(0) -{c_i \over \omega^2_i} x(0)\right) \cos(\omega_i t) \cr\cr
&+&  {c_i\over \omega_i} p_i(0) \sin(\omega_i t) \bigg\} .   \label{f}
\ea
It should be noted that the internal noise has the usual expression when the system is not influenced by an external force~\cite{Zwanzig2001}. It is a stochastic quantity determined by the random initial conditions $(x(0), p(0) ,x_{i}(0)$, $p_{i}(0))$. We can see that the conservative electric force ${\mathcal F}_e(t)=QE(t)$ in Eq.~\ref{gle1} comes directly from the derivative of the electric potential energy $U_e(x,t)=-QE(t) x$; however, the induced force ${\mathcal F}_{ein}(t)$ is not derived from a set of ``electric potentials'' $\bar U_{ei}(x_i,t)=-q_iE(t) x_i$, but rather it is an induced force coming from the interaction between the electric field and bath particles. In practically all studies on the fluctuation theorems reported in the literature, the applied force driving the system out of equilibrium is treated as a deterministic time-dependent protocol~\cite{Bochkov1981}-\cite{Rao2018}. Here, our time-dependent protocol playing the role of an effective deterministic electric force is the sum $F_e(t)={\mathcal F}_e(t)+{\mathcal F}_{in}(t)$. In the following, we will prove the validity of the DFT for the total entropy production for such a protocol.

\subsection{Kubos's second FDT}

Once we have identified the effective electric force $F_e(t)$ as a deterministic one, the GLE (\ref{gle1}) can be written as (for simplicity, we take $m=1$)  
\be
\dot v(t)=-{\mathcal V}_0^{\prime}(x) -\int_0^t \gamma(t-t^{\prime})  \, v(t^{\prime}) \, dt^{\prime} + F_e(t) +\eta(t) . \label{gle2} 
\ee
We now follow the same Zwanzig's methodology~\cite{Zwanzig2001}, to prove that the noise correlation function satisfies Kubo's second FDT, that is, 
\be
\langle \eta(t) \eta(t^{\prime})\rangle
=k_BT\gamma(t-t^{\prime}) , \label{Kfdt} 
\ee
which guarantees that the GLE in Eq.~\ref{gle2} without the effective force is stationary in the long-time limit.

\subsection{Modified Kubo's second FDT}

It should be noted that the theoretical model proposed in this work is the same as that reported in Ref.~\cite{Cui2018}, in which a modified Kubo's second FDT has been obtained for an oscillating electric field (see Eq. [14] in Ref.~\cite{Cui2018}). However, the modified Kubo's FDT can also be formulated for an arbitrary but well-behaved time-dependent electric field. In the particular case of an oscillating electric field, the modified theorem reduces to the same result reported in Ref.~\cite{Cui2018}. To achieve the goal, we use the following strategy: we define the effective noise as $\eta_{\text{eff}}(t)=\eta(t)+{\mathcal F}_{ein}(t)$, such that $\langle \eta_{\text{eff}}(t)\rangle={\mathcal F}_{ein}(t)$, because $\langle \eta(t)\rangle=0$, and therefore the GLE (\ref{gle2}) is now written as 
\be
\ddot x(t)=-{\mathcal V}_0^{\prime}(x)-\int_0^t \gamma(t-t^{\prime})  \, \dot x(t^{\prime}) \, dt^{\prime} + {\mathcal F}_e(t)+\eta_{\text{eff}}(t), \label{cgle1} 
\ee
Following the same algebraic steps given in \cite{Cui2018}, it is easy to show that the FDT becomes
\begin{equation}
\langle \eta_{\text{eff}}(t) \eta_{\text{eff}}(t^{\prime})\rangle=k_BT\gamma(t-t^{\prime})+  {\mathcal F}_{ein}(t){\mathcal F}_{ein}(t^{\prime}),  \label{Cmfdt} 
\end{equation}
which in principle would be the modified Kubo's second FDT. In the particular case of an oscillating electric field $E(t)=E_0\sin(\Omega t)$ and assuming the condition 
$\omega_i\gg \Omega$, it has been shown that ${\mathcal F}_{ein}(t)={\mathcal C} Q E(t)$, with ${\mathcal C}=\sum_i c_i/\omega_i^2$ and therefore~\cite{Cui2018} 
\begin{equation}
\langle \eta_{\text{eff}}(t) \eta_{\text{eff}}(t^{\prime})\rangle=k_BT\gamma(t-t^{\prime})+ (\mathcal C Q)^2 E(t)E(t^{\prime}),  \label{Cfdt} 
\end{equation}
On the other hand, in our approach to demonstrate the DFT for the total entropy production, the GPD $P(x,t)$ along a single stochastic trajectory is necessary. This requires an expression of the first two moments, $\langle x(t)\rangle$ and $\langle x^2(t)\rangle$, derived from the solution of the GLE in Eq.~\ref{gle1} for the harmonic potential trap ${\mathcal V}_0(x)=(k/2)x^2$. Under these conditions, the following question then arises: Which of the two FDTs (Eq.~\ref{Kfdt}) or (Eq.~\ref{Cmfdt}) should be used to obtain the correct Gaussian probability density? In the next sections, we show that both FDTs lead to the same GPD.   


\section{Solution of the GLE and GPD} \label{sec:3}

In this section, we solve the GLE and use both Kubo's second FDT and its modified version to calculate the GPD. In one case, the GLE of a Brownian particle bound to a harmonic potential trap and under the action of an effective electric force $F_e(t)$ is explicitly solved, and Kubo's second FDT is used to calculate the GPD. In the other case, the GLE is solved by considering an effective noise $\eta_{\text{eff}}(t)$; then the modified Kubo's second FDT is used to obtain the same aforementioned GPD.

\subsection{Solution of the GLE considering the effective force $F_e(t)$}

As discussed above, for the proposed model in this work, we have defined the effective force $F_{e}(t)=\mathcal{F}_{e}(t)+{\mathcal F}_{ein}(t)$.  For a harmonic potential centered at the origin (because $m=1$, then $\omega^2=k$), the GLE in Eq.~\ref{gle1} reads
\be 
\ddot x(t)=-\int_0^t \gamma(t-t^{\prime})  \, \dot x(t^{\prime}) \, dt^{\prime} -k x + F_e(t)+\eta(t).  \label{gle3} 
\ee
We now define the new variables $X=x-\langle x\rangle$, $V=v-\langle v\rangle$, to get
\ba
\ddot X+ k X +\int_0^t\gamma(t-t')\,\dot X(t')\, dt' &=&\eta(t) \label{gleX}, \\
 \langle \ddot x\rangle  + k \langle x\rangle 
+\int_0^t\gamma(t-t') \langle \dot x(t') \rangle\, dt'&=&F_e(t) . 
 \label{mglex}  \ea
The solution of Eqs.~\ref{gle2},~\ref{gleX} and~\ref{mglex} can be obtained using the Laplace transforms, yielding to 
\ba 
x(t)&=&x_0 H(t)+v_0 G(t) + \int_0^t G(t-t')\, F_e(t') dt' \cr\cr
&+&\int_0^t G(t-t') \,\eta(t') dt' , \label{solx} \\
X(t)&=&X_0 H(t)+V_0 G(t) +\int_0^t G(t-t') \, \eta(t') \,dt' ,  ~~\label{solX} ~~  \\
\langle x(t)\rangle&=&
\langle x_0\rangle H(t)+\langle v_0\rangle G(t) 
+ \int_0^t G(t-t')\, F_e(t') dt' , \nonumber\\
 \label{solmx} \ea
being $x_0=x(0)$, $v_0=v(0)$, $X_0=X(0)$, $V_0=V(0)$, the initial conditions at time $t=0$.   The  functions $H(t)$ and $G(t)$ are defined by 
\be
H(t)=1-k\int_0^t G(t') \,dt'    \label{chi0}\, , \ee
with $G(t)$ the inverse Laplace transform of $\hat G(s)$, that is, 
$G(t)={\mathcal L}^{-1}\{\hat G(s)\}$, such that 
\be \label{hatG}
\hat G(s)={1\over s^2+s\hat\gamma(s)+k} ,  \ee
and $\hat\gamma(s)$ is the Laplace transform of $\gamma(t)$, i.e. 
$\hat\gamma(s)=\int_0^{\infty} e^{-st} \gamma(t) dt$. On the other hand, 
it is easy to see from Eq. (\ref{solx}) that at time $t=0$,  
$x_0=x_0 H(0)+v_o G(0)$, from which we can conclude that $H(0)=1$ and $G(0)=0$.



\subsection{GPD and Kubo's second FDT}

Because $x(t)$ given in Eq.~\ref{solx} is a linear function of the Gaussian colored noise $\eta(t)$, it is also a Gaussian stochastic process with a probability density 
\be \label{pxt}
P(x,\tau)=\sqrt{k\over 2\pi \sigma^2_x(\tau)}\, \exp\left[- {[x-\langle x(\tau)\rangle]^2\over 2 \sigma^2_x(\tau)} \right]. \ee
where 
\be \label{solmx2} 
\langle x(\tau)\rangle=
 \int_0^{\tau}  G(\tau-t)\, F_e(t) \, dt ,
 \ee
if $\langle x_0\rangle=\langle v_0\rangle=0$ is assumed. It should be noted that the variance $\sigma^2_x(\tau)=\langle (x-\langle x\rangle)^2=\langle X^2(\tau)\rangle=\sigma^2_X(\tau)$, and also 
$\langle X_0\rangle=\langle V_0\rangle=0$. On the other side, by assuming that $\langle X_0 V_0\rangle=\langle X_0 \eta(t')\rangle=\langle V_0\eta(t')\rangle=0$, it can be shown from Eq. (\ref{solX}) that
\ba  
&&\langle X^2(\tau)\rangle=\langle X^2_0\rangle H^2(\tau)+\langle V^2_0\rangle G^2(\tau)\cr\cr
&+&\int_0^{\tau}\int_0^{\tau} G(\tau-t_1) G(\tau-t_2) \langle \eta(t_1)\eta(t_2)\rangle dt_1 dt_2 .~~~ \label{X2a}
\ea 
On the other side, if $X_0$ satisfies the equilibrium canonical distribution defined as $P(X_0)=\sqrt{k/2\pi\, k_BT}\exp(-k X^2_0/2 k_BT)$, and $V_0$ the Maxwellian distribution, $P(V_0)=\sqrt{1/2\pi\, k_BT}\exp(- V^2_0/2 k_BT)$, then $\langle X_0^2\rangle=k_BT/k$ and $\langle V^2_0\rangle=k_BT$. Next we will show that the variance $\sigma^2_X(\tau)=\langle X^2(\tau)\rangle$ is the same as $\langle X^2_0\rangle$. To prove this, we calculate the double integral of (\ref{X2a}) using Kubo's second FDT (\ref{Kfdt}) and proceed as follows: we define  
\ba 
&&R(\tau)=\int_0^{\tau}\int_0^{\tau} G(\tau-t_1) G(\tau-t_2) \langle \eta(t_1)\eta(t_2)\rangle dt_1dt_2\cr\cr
&=&k_BT\int_0^{\tau}\int_0^{\tau}G(\tau-t_1) G(\tau-t_2)\gamma(t_1-t_2) \, dt_1dt_2\cr\cr
&=&2k_BT\int_0^{\tau} G(t)\, dt\int_0^t G(t_1)\gamma (t-t_1)\, dt_1. \label{Ra}
\ea
Taking the time derivative of $R(\tau)$ we show that
\ba 
\dot R(\tau)&=&2k_B T G(\tau)\int_0^{\tau} G(t_1)\gamma (\tau-t_1)\, dt_1 \cr\cr
&=& 2k_BT G(\tau){\mathcal L}^{-1}\{ \hat\gamma(s)\hat G(s)\} .\label{dR}
\ea
Using Eq. (\ref{hatG}), it is possible to show that  
\be \label{Rb} 
R(\tau)=k_B T\bigg\{\frac{1}{k}[1-H^2(\tau)]- G^2(\tau) \bigg\}, \ee
and therefore
\ba
\langle X^2(\tau)\rangle
&=&k_BT\bigg\{ {1\over k} H^2(\tau)+G^2(\tau)\cr\cr
&+&\left. {1\over k}[1-H^2(\tau)]-G^2(\tau) \right\} ={k_B T\over k}, \label{X2b} \ea
showing that $\sigma_X^2(\tau)= k_BT/k=\langle X^2_0\rangle$. Moreover  the GPD (\ref{pxt}) becomes 
\begin{align}\label{pdxt}
   P(x,\tau)= {\sqrt{k\over 2\pi\, k_BT}} \exp\left [-{k(x-\langle x\rangle)^2\over 2 k_B T}  \right]. 
\end{align}
Clearly $P(x,\tau)=P(X,\tau)$, because $X=x-\langle x\rangle$.
Also, for all time the variance $\sigma^2_x(\tau)=\langle X_0^2\rangle=k_BT/k$. In the absence of the external electric field, $F_e(t)=0$, also $\langle x(\tau)\rangle=0$, and therefore $P(x,\tau)=P(X_0)$, as expected.



\subsection{Solution of the GLE considering an effective noise $\eta_{\text{eff}}(t)$}

We now proceed to obtain the GPD using the modified Kubo's second FDT given by Eq.~\ref{Cmfdt}. To achieve the goal, we begin with the GLE in Eq.~\ref{cgle1} for a harmonic potential, that is
\be
\ddot x(t)=-\int_0^t \gamma(t-t^{\prime})  \, \dot x(t^{\prime}) \, dt^{\prime} -k x + {\mathcal F}_e(t)+\eta_{\text{eff}}(t). \label{cgle2} 
\ee
The solution can be written as   
\ba 
x(t)&=&x_0 H(t)+v_0 G(t) + \int_0^t G(t-t')\, {\mathcal F}_e(t') dt' \cr\cr
&+&\int_0^t G(t-t') \,\eta_{\text{eff}}(t') dt' . \label{csolx} \ea
Because $F_e(t)={\mathcal F}_e(t) + \langle \eta_{\text{eff}}(t)\rangle={\mathcal F}_e(t)+{\mathcal F}_{ein}(t)$, and assuming that $\langle x_0\rangle=0$, $\langle v_0\rangle=0$, $\langle x_0\eta(t)\rangle=0$ and $\langle v_0\eta(t)\rangle=0$, then $\langle x_0\eta_{\text{eff}}(t)\rangle=\langle x_0\rangle{\mathcal F}_{ein}(t)+ \langle x_0\eta(t)\rangle=0$, and also $\langle v_0\eta_{\text{eff}}(t)\rangle=0$, the average value $\langle x(t)\rangle$ now reads 
\ba
\langle x(t)\rangle&=&
\langle x_0\rangle H(t)+\langle v_0\rangle G(t) + \int_0^t G(t-t')\, F_e(t') dt' \cr\cr
&=&\int_0^t G(t-t')\, F_e(t') dt', 
 \label{csolmx} \ea
which is the same as the one given in Eq.~\ref{solmx2}. The reason for this equality is that the average value of the solution \ref{solx} is the same as the average value of Eq.~ \ref{csolx}.

\subsection{GPD and the modified Kubo's second FDT}

Again, to obtain the variance $\sigma_x^2(t)$ we need the second moment of $x(t)$, which is obtained from the solution in Eq.~\ref{csolx}, giving as a result 
\ba 
&&\langle x^2(t)\rangle=\langle x^2_0\rangle H^2(t)+\langle v^2_0\rangle G^2(t) \cr\cr
&*& \int_0^t\int_0^t G(t-t_1')G(t-t_2^{\prime}) \, {\mathcal F}_e(t_1') {\mathcal F}_e(t_2^{\prime}) dt_1' dt_2'  \cr\cr
&+&2\int_0^t\int_0^t G(t-t_1') G(t-t_2') \,\langle {\mathcal F}_e(t_1')\eta_{\text{eff}} (t_2')\rangle dt_1' dt_2' \cr\cr
&+&\int_0^t\int_0^t G(t-t_1') G(t-t_2') \,\langle \eta_{\text{eff}}(t_1')\eta_{\text{eff}}(t_2')\rangle dt_1' dt_2'. \nonumber \\
 \label{csm1} \ea
However, the effective noise correlation function $\langle \eta_{\text{eff}}(t_1')\eta_{\text{eff}}(t_2')\rangle$ is given by Eq.~\ref{Cmfdt}, and given that $\eta_{\text{eff}}(t)={\mathcal F}_{ein}(t)+\eta(t)$, for which $\langle {\mathcal F}_e(t_1')\eta_{\text{eff}} (t_2')\rangle={\mathcal F}_e(t_1'){\mathcal F}_e(t_2')$ one gets
\ba 
&&\langle x^2(t)\rangle=\langle x^2_0\rangle H^2(t)+\langle v^2_0\rangle G^2(t) \cr\cr
&+& \int_0^t\int_0^t G(t-t_1')G(t-t_2^{\prime}) \, {\mathcal F}_e(t_1') {\mathcal F}_e(t_2^{\prime}) dt_1' dt_2'  \cr\cr
&+&2\int_0^t\int_0^t G(t-t_1') G(t-t_2') \,{\mathcal F}_e(t_1'){\mathcal F}_{ein}(t_2') dt_1' dt_2' \cr\cr
&+&\int_0^t\int_0^t G(t-t_1') G(t-t_2') \, {\mathcal F}_{ein}(t_1'){\mathcal F}_{ein}(t_2')  dt_1' dt_2'\cr\cr
&+&k_BT\int_0^t\int_0^t G(t-t_1') G(t-t_2') \, \gamma(t_2'-t_1')\, dt_1' dt_2'.
\nonumber \\
 \label{csm2} \ea
However, the sums of the third, fourth, and fifth terms are equal to $\langle x(t)\rangle ^2$, and the last one is the same as that in Eq.~\ref{Rb}, previously obtained in Sec.~\label{sec:3} B. Also, $\langle x_0^2\rangle=\langle X_0^2\rangle=k_BT/k$, $\langle v_0^2\rangle=\langle V_0^2\rangle=k_BT$, and therefore, the variance at time $t=\tau$ becomes
\ba
&&\sigma_x^2(\tau)=k_BT\left[
{1\over k}H^2(\tau)+G^2(\tau)\right.\cr\cr
&+&\left.{1\over k}[1-H^2(\tau)]-G^2(\tau)\right] ={k_B T\over k}, \label{cvarx} \ea
which is the expected result. According to the mean value (Eq.~\ref{csolmx}) and the variance (Eq.~\ref{cvarx}), we obtain the same GPD as given by Eq. (\ref{pdxt}). According to the results obtained in this section, we can conclude that even when the bath-particle system is influenced by any well-behaved time-dependent force, such as an external electric field, Kubo's second FDT remains unchanged.


\section{DFT for the total entropy production} 
\label{sec:4}

In this section, we demonstrate the validity of the detailed fluctuation theorem for the total entropy production for two protocols of the external time-dependent force. The first one is a general time-dependent external force applied to the system under a harmonic potential trap, and the second is the translation of the center of the harmonic potential trap. We employ the GLE given by Eq.~\ref{gle3}, as $F_e(t)$ is the deterministic force responsible for driving the system out of equilibrium. So that, according to stochastic thermodynamics~ \cite{Seifert2005,Ohkuma2007,Sekimoto2010}, the first law-like balance between the applied work $W$, the change in internal energy $\Delta U$, and the dissipated heat $Q$ to the bath can be calculated along a single stochastic trajectory $x(t)$ over a finite time interval $\tau$. This first-law-like reads
\be \label{fl}
Q=W-\Delta U,
\ee
where the work $W\equiv W(\tau)$ is defined as 
\be
W=\int_0^{\tau} {\partial {\mathcal U}(x,\lambda(t))\over\partial \lambda} \dot\lambda\, \, dt ,
\ee
with ${\mathcal U}(x,\lambda(t))$ an effective external potential energy and $\lambda(t)$ an external parameter changing under a time-dependent protocol. As usual, the total entropy production $\Delta s_{tot}$ is the sum of the change in the medium entropy  over the time interval, $\Delta s_m=Q/T$, and the nonequilibrium Gibbs entropy $S$ of the system is defined as~\cite{Seifert2005}
\be \label{S}
S(t)=- k_B\int P(x,t) \ln P(x,t) \, dx, 
=\langle s(t)\rangle, \ee
where $P(x,t)$ is the probability density as before. This definition suggests the definition of a trajectory-dependent entropy for the particle as
\be \label{s}
s(t)=- k_B\ln P(x(t),t).
\ee
where $P(x, t)$ is the solution of the Fokker-Planck equation, evaluated along the stochastic trajectory. For a given trajectory $x(t)$, the entropy $s(t)$ depends on the probability density of the particle at the initial time $t=0$, $P(x_0)=P(x_0,0)$, and
thus contains information about the whole ensemble. The change in the system entropy for any trajectory of duration $\tau$ reads
\be \label{Ds1}
\Delta s=-k_B \ln\left[ P(x,\tau)\over P(x_0)\right] .
\ee
Therefore, the change in total entropy production along a trajectory over a finite time interval $\tau$ is~\cite{Seifert2005}
\be \label{Dst}
\Delta s_{tot}= \Delta s_m +\Delta s. 
\ee
Using this definition, the IFT, $\langle e^{-\Delta s_m}\rangle=1$ can be derived~\cite{Seifert2005}, where the angular brackets denote an average over the statistical ensemble of realizations or over the ensemble of finite time trajectories. Also, in the nonequilibrium steady state over a finite time interval, the DFT holds. So, the change in the total entropy production thus becomes~\cite{Seifert2005}  
\be \label{Dst1}
\Delta s_{tot}={W-\Delta U\over T} 
-k_B\ln\left[{P(x,\tau)\over P(x_0)}\right]. 
\ee
Here, for the harmonic trap, we assume that the system is initially prepared in thermal equilibrium in such a way that $P(x_0)$ satisfies the canonical distribution given by
\be \label{px0}
P(x_0)=\sqrt{k\over 2\pi k_BT}\, \exp\left(- {k x_0^2\over 2k_BT} \right), \ee
and $P(x,t)$ is the same as Eq.~\ref{pdxt}. Next, because the $\Delta s_{tot}$ depends on the thermodynamic work $W$, we proceed to calculate the work statistics for the two proposed models.

\subsection{Influence of an effective time-dependent external force over the harmonic trap}

In this case, the effective potential is given by ${\mathcal U}(x, \lambda(t))=(k/2)x^2- F_e(t) x$, and $\lambda(t)\equiv F_e(t))$. The total work done on the system over a finite time $\tau$ becomes 
\be \label{Wt}
W=- \int_0^{\tau} \dot F_e(t)\, x(t) \,dt , \ee
and the change in the internal energy is
\ba \label{DU}
\Delta U &=& U(x(\tau),\tau)-U(x_0,0)\cr\cr
&=&  {1\over 2} k \left(x^2-x_{0}^{2}\right)- x F_e(\tau),
\ea
where for simplicity we assume that $F_e(0)=QE(0)=0$, as by definition ${\mathcal F}(0)=0$. Because $\langle x_0\rangle=0$ and $\langle v_0\rangle=0$, we thus have that
\ba \label{mWt}
\langle W\rangle&=&- \int_0^{\tau} \dot F_e(t) \langle x(t)\rangle\, dt \cr\cr
&=&-\int_0^{\tau} dt \int_0^t dt' \,  \dot F_e(t) \,G(t-t') \, F_e(t') . \ea
Upon substitution of $x=X+\langle x\rangle$ into Eq.~\ref{Wt}, it is easy to show that the work variance $\sigma^2_W=\langle W^2\rangle-\langle W \rangle^2$ can be written as  
\ba \label{varW1}
\sigma^2_W&=&\int_0^{\tau}\int_0^{\tau} \langle X(t)X(t')\rangle \dot F_e(t) \dot F_e(t')\, dt dt' \cr\cr
&=&2 \int_0^{\tau} dt\, \dot F_e(t) \int_0^t \langle X(t)X(t')\rangle \dot F_e(t') \, dt' .
 \ea
In Ref~\cite{Jimenez2015} it has been shown that the process $X(t)$ is stationary, and  thus $\langle X(t)X(t')\rangle=\langle X(0)X(t-t')\rangle$ and also $\langle X_0^2\rangle=k_BT/k$. Using the solution given by Eq. (\ref{solX}) and assuming that $\langle X_0 \eta(t)\rangle=0$, we obtain 
\be \label{cfX2}
\langle X_0 X(t-t')\rangle=\langle X_0^2\rangle H(t-t')+\langle X_0 V_0\rangle G(t-t'), \ee
but also $\langle X_0 V_0\rangle=0$, and therefore 
\be \label{corX3}
\langle X_0 X(t-t')\rangle= {k_BT\over k}\, H(t-t'). \ee
So, the work variance in Eq.~\ref{varW1} can be written as 
\be \label{varW2}
\sigma^2_W= {2 k_BT\over k}\int_0^{\tau} dt\, \dot F_e(t) \int_0^t H(t-t')\dot F_e(t') \, dt' .
\ee
Integrating by parts the integral inside, we conclude that 
\ba \label{varW3}
\sigma^2_W&=&{2 k_BT\over k} \int_0^{\tau} dt\, \dot F_e(t) \bigg[ F_e(t) H(0) \cr\cr
&-&k \int_0^t G(t-t') F_e(t') \,dt'\bigg] \cr\cr
&=&{2 k_BT\over k} \bigg[ - \int_0^{\tau} \dot F_e(t)\langle x(t)\rangle \,dt + {F_e^2(\tau)\over 2 k} \bigg] \cr\cr
&=&2 k_BT\left[\langle W\rangle + {F_e^2(\tau)\over 2 k} \right], \ea
Once this is done, we use Eqs.~\ref{pdxt},~\ref{Dst1}, and after some algebra, we get  
\be \label{Dsta} 
	\Delta s_{tot}=\frac{1}{T}\left(W +x F_e - k x\langle x\rangle +\frac{1}{2}k\langle x\rangle^{2} \right).
\ee
It can be seen from Eqs.~\ref{Wt} and~\ref{Dsta} that $W$ as well as $\Delta s_{tot}$ are linear functions of $x$, and this in turn is also a linear function of the Gaussian noise $\eta(t)$. This means that $\Delta s_{tot}$ is a Gaussian stochastic process and its probability density is given by 
\ba
P(\Delta s_{tot})={1\over\sqrt{2\pi\, \sigma_s^2(t)}} \exp\left[-{(\Delta s_{tot}-  \langle \Delta s_{tot}\rangle)^2 \over 2\sigma_s^2(t)} \right],~~ \label{eppd}
\ea
where $\langle \Delta s_{tot}\rangle$ and $\sigma_s^2(t)=\langle (\Delta s_{tot})^2\rangle- \langle \Delta s_{tot}\rangle^2$ are the mean value and variance of the total entropy production. From Eq.~\ref{Dsta}, it is easy to see that 
\be \label{mDst}
	\langle \Delta s_{tot}\rangle={1\over T} \left(\langle W\rangle - {1\over 2}k \langle x\rangle^2 +\langle x \rangle F_e\right) .
\ee
After some algebra, the variance can be written as   
\ba \label{varsta}
\sigma^2_s(\tau)&=& {1\over T} \left( {\sigma_W^2\over T}+{F_e^2\over k} +   
k \langle x\rangle^2 -2\langle x\rangle F_e\right) \cr\cr
&+&{1\over T^2}[\langle W x\rangle- \langle W\rangle\langle x\rangle][2F_e-2k\langle x\rangle] .
\ea
 If the Boltzmann constant $k_B$ is absorbed into the temperature $T$ of Eq.~\ref{varW3}, we have that  
\ba \label{varsta}
\sigma^2_s(\tau)&=& {1\over T} \left( {\sigma_W^2\over T}+{F_e^2\over k} +   
k \langle x\rangle^2 -2\langle x\rangle F_e\right) \cr\cr
&+&{1\over T^2}[\langle W x\rangle- \langle W\rangle\langle x\rangle][2F_e-2k\langle x\rangle] \cr\cr
&=&{1\over T} \left( 2\langle W\rangle+{2F_e^2\over k} +   
k \langle x\rangle^2 -2\langle x\rangle F_e\right) \cr\cr
&+&{2\over T^2}[\langle W x\rangle- \langle W\rangle\langle x\rangle][F_e-k\langle x\rangle] ,
\ea
and according to Eq.~\ref{a2} given in App.~\ref{app:a}, we show that    
\ba \label{varps}
\sigma^2_s(\tau)&=& {2\over T}\left(\langle W\rangle - {1\over 2}k \langle x\rangle^2 +\langle x \rangle F_e\right) 
\cr\cr
&= & 2\langle \Delta s_{tot} \rangle  .
\ea
Under these conditions, the DFT for the total entropy production is then 
\be
{P(\Delta s_{tot})\over P(-\Delta s_{tot})} = e^{\Delta s_{tot}} . \label{ftep1}
\ee
From this DFT for the total entropy production, we immediately conclude that the IFT also satisfies $\langle e^{-\Delta s_{tot}}\rangle=1$, because  $\langle e^{-\Delta s_{tot}}\rangle=\int e^{-\Delta s_{tot}} P(\Delta s_{tot}) \,d\Delta s_{tot}=\int P(-\Delta s_{tot}) \,d\Delta s_{tot}=1$, since  $P(-\Delta s_{tot})$ is normalized.

\subsection{Dragging of the center of the harmonic potential trap}

The other solvable model we consider is the dragging of the harmonic potential’s center with arbitrary velocity during a finite time interval. In this case, the effective potential thus becomes $\hat{\mathcal U}(x,\lambda(t))=(k/2)\left(x-F_e(t)/k\right)^2$, where the time-dependent protocol reads $\lambda(t)\equiv F_e(t)/ k$, representing the position of the harmonic potential minimum. It is assumed that at time $t=0$, the potential minimum is located in the origin of coordinates, $\lambda(0)=F_e(0)/k=0$, and $\dot\lambda(t)=\dot F_e(t)/k$, is the velocity for which the potential minimum is arbitrarily dragged. For this second model, the corresponding thermodynamic work reads 
\be \label{hWt}
\hat W=- \int_0^{\tau} \dot F_e(t) x(t) \,dt
+{F_e^2(\tau)\over 2k}. \ee
and the mean work is 
\ba
&&\langle\hat W\rangle=- \int_0^{\tau} \dot F_e(t) \langle x(t)\rangle\, dt +{F_e^2(\tau)\over 2k} \cr\cr
&=&-\int_0^{\tau} dt \int_0^t dt' \,  \dot F_e(t) G(t-t') \, F_e(t') +{F_e^2(\tau)\over 2k} . ~~
\label{mhW}\ea
Substituting $x=X+\langle x\rangle$ into Eq.~\ref{hWt} we obtain  
\ba \label{hWt2}
\hat W&=&- \int_0^{\tau} \dot F_e(t) X(t) \,dt
- \int_0^{\tau} \dot F_e(t) \langle x(t)\rangle \, dt 
+
{F_e^2(\tau)\over 2k} \cr\cr
& = &- \int_0^{\tau} \dot F_e (t) X(t) \,dt+ \langle \hat W\rangle. \ea
In a similar way as before, the work variance thus becomes
\ba \label{hvar}
&&\sigma^2_{\hat W} =2\int_0^{\tau}dt\, \dot F_e(t) \int_0^t \langle X(t)X(t')\rangle\, \dot F_e(t') \,dt \cr\cr
&=& {2 k_BT\over k} \int_0^{\tau} dt\, \dot F_e(t) \int_0^t H(t-t')\dot F_e(t') \, dt'\cr\cr
&=&{2 k_BT\over k} \bigg[ - \int_0^{\tau}dt \int_0^t dt'\, \dot F_e(t) G(t-t') F_e(t') + {F_e^2(\tau)\over 2 k} \bigg]
. \cr\cr
&=& {2 k_BT\over k} \langle \hat W \rangle. \ea
The change in the internal energy during a time $\tau$ now becomes 
\be \label{DU2}
\Delta U=  {1\over 2} k\left(x- {F_e(\tau)\over k}\right) -{k\over 2} x_0^2,
\ee
and the change in the total entropy production reads
\be \label{Dsta2}
	\Delta \hat s_{tot}={\hat W\over T} -{F_e^2\over 2kT}+{x F_e\over T} - {k x\langle x\rangle\over T} +{k\langle x\rangle^2\over 2T} ,
\ee
and therefore $P(\Delta \hat s_{tot})$ is again a Gaussian distribution. Following similar algebraic steps as before, we can conclude that 
\be \label{mDst2}
	\langle \Delta \hat s_{tot}\rangle={1\over T} \left(\langle \hat W\rangle - {F_e^2\over 2k}- {k\langle x\rangle^2\over 2} +\langle x \rangle F_e\right) ,
\ee
and the variance satisfies 
\ba \label{varsta}
\sigma^2_s(\tau)&=& {1\over T} \left(2\langle \hat W\rangle - {F_e^2\over k}- k\langle x\rangle^2 +2\langle x \rangle F_e\right) \cr\cr
&=& 2\langle \Delta \hat s_{tot}\rangle.
\ea
Therefore, the DFT as well as the IFT for the total entropy production hold. 



\begin{figure*}[t]
\centering
\includegraphics[width=\textwidth]{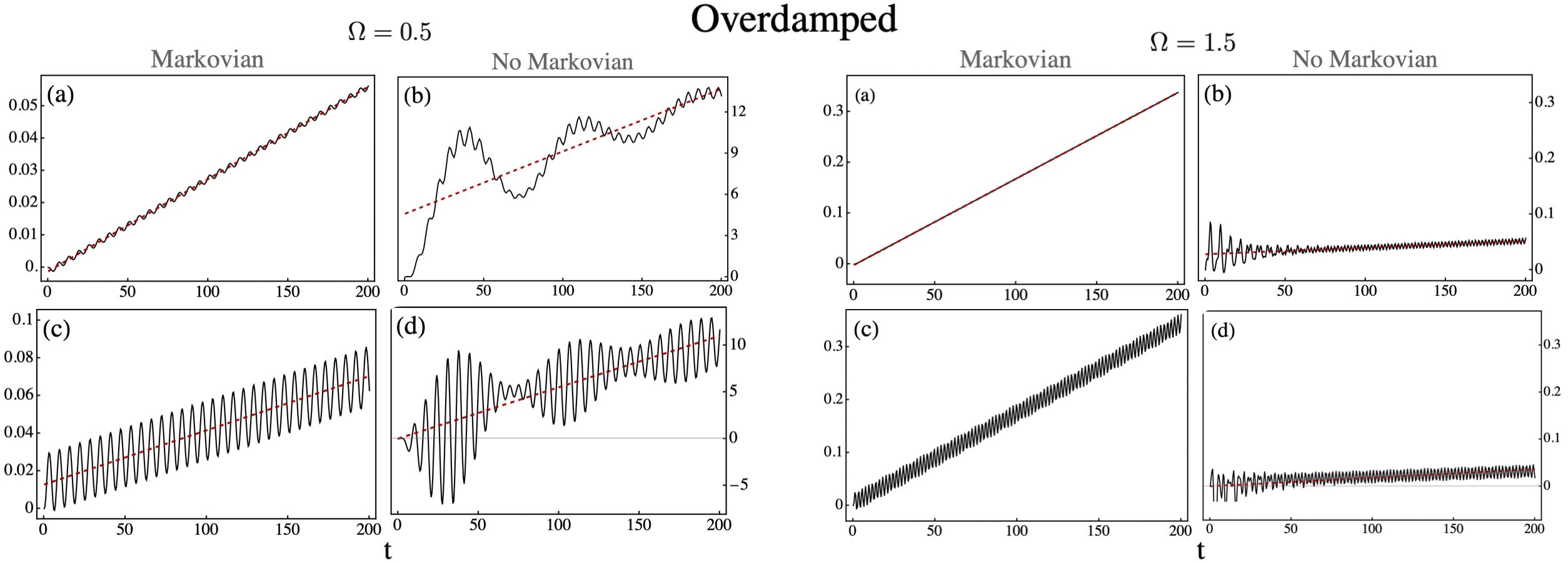}
\caption{Average work and total entropy production in the Markovian and non-Markovian regimes in the overdamped limit. The figure is arranged in two sets corresponding to different driving frequencies. Panels (a–d) show results for $\Omega=0.5$, with parameters  $\gamma_0 =10$, $\omega = 4$, and correlation time $\tau_{c}=5$, chosen such that $\omega \gg \Omega$ and $\gamma_0 > 2\omega$. Panels (a-b) display the average work, while (c-d) show the total entropy production, for the Markovian (left) and non-Markovian (right) cases. Panels (e–h) correspond to $\Omega=1.5$, where (e-f) present the average work and (g-h) the total entropy production for the same ordering.}
\label{fig:1}
\end{figure*}

\begin{figure*}[t]
\centering
    \includegraphics[width=\textwidth]{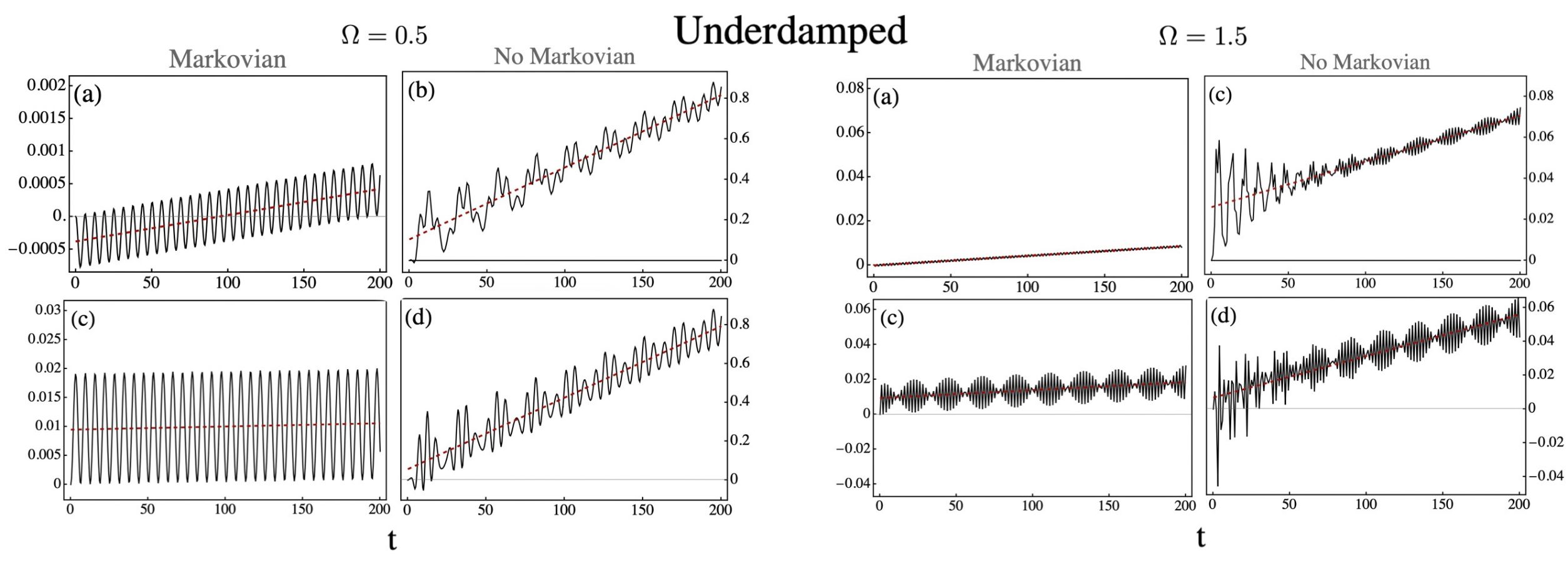}
\caption{Average work and total entropy production in the Markovian and non-Markovian regimes in the underdamped limit. The figure is arranged in two sets corresponding to different driving frequencies. Panels (a–d) show results for $\Omega=0.5$, with parameters  $\gamma_0= 0.5$  and $\omega = 5$ and correlation time $\tau_{c}=5$, $\omega > 2\gamma_0$, chosen such that $\omega \gg \Omega$ and $\omega > 2\gamma_0$. Panels (a-b) display the average work, while (c-d) show the total entropy production, for the Markovian (left) and non-Markovian (right) cases. Panels (e–h) correspond to $\Omega=1.5$, where (e-f) present the average work and (g-h) the total entropy production for the same ordering.}
\label{fig:2}
\end{figure*} 

\section{Average of the total entropy production}
\label{sec:5}

Here, we obtain explicit results for the average of the total entropy production in Eq.~\ref{mDst}, which in turn is a function of $\langle x(t)\rangle$ and  $\langle W(t)\rangle$. In the following, we will consider a particular case of the time-dependent external force studied in Sec.~\ref {sec:4} A, for which the mean value of the position and work are given by Eqs.~\ref{csolmx} and~\ref{mWt}, respectively, and a specific memory kernel. We study an oscillating electric field $E(t)=E_0\sin(\Omega t)$ and an Ornstein-Uhlenbeck type friction memory kernel. For an oscillating electric field, the induced force given by Eq.~\ref{F} after integration reads as 
 \be \label{eqa}
{\mathcal F}_{ein}(t)=\sum_i c_i q_i E_0 \frac{ [\Omega  \sin(\omega_i t) - \omega_i \sin(\Omega t)] }
{\omega_i(\Omega^2-\omega^2_i)} .
\ee
However, it has been shown in Ref.~\cite{Cui2018} that in the case for which $\omega_i\gg \Omega$, it can be approximated by   
 \be \label{eqb}
{\mathcal F}_{ein}(t) \approx\sum_i {c_i q_i\over\omega_i^2} E_0 \sin(\Omega t) = 
{\mathcal C} \, QE(t)\ee
where we assume that $q_i=Q$, and  ${\mathcal C}=\sum_i(c_i/\omega_i^2)$ is a dimensionless constant expressed in terms of the coupling parameter $c_i$. 

Therefore, the effective electric force becomes 
 \ba \label{Fe-a}
F_e(t)&=&{\mathcal F}_e(t)+{\mathcal F}_{ein}(t)=Q E(t)+{\mathcal C}QE(t)\cr
&=&{\mathcal K} Q E(t)={\mathcal K}Q E_0\sin(\Omega t),
\ea
where ${\mathcal K}=1+{\mathcal C}$, which clearly shows that the effective force is proportional to the applied electric field responsible for driving the system out of equilibrium.

On the other hand, to show that the friction memory kernel $\gamma(t)$ given by Eq.~\ref{g} is of the OU type, we proceed as follows: in the thermodynamic limit, the memory function should tend to a decaying function of time. The form of this function depends on the so-called spectral density of the bath particles, which is given by~\cite{Luczka2005}
 \begin{equation}
 \rho(\omega)=\sum_i {m_ic^2_i\over \omega_i} \delta(\omega-\omega_i) ,
  \label{ro}
\end{equation}
which allows one to write the memory function (Eq.~\ref{g}) as
 \begin{equation}
 \gamma(t) =\int_0^{\infty} {\rho(\omega)\over\omega} \cos(\omega) \, d\omega . \label{gama}
\end{equation}
If we propose $\rho(\omega)= (2\gamma_0 / \pi\tau^2) (\omega/ \omega^2+\tau_c^{-2})$, we have
 \begin{equation}
 \gamma(t)= {2\gamma_0 \over \pi \tau_c^2}   
 \int_0^{\infty}  { \cos(\omega t) \over \omega^2+\tau_c^{-2}} \, d\omega = 
 {\gamma_0 \over \tau_c}  e^{-t/\tau_c} ,  \label{gama2}
\end{equation}
where $\tau_{c}$ is the correlation time. Therefore, the OU-type friction-memory kernel reads
 \begin{equation}
\gamma(t-t')=   {\gamma_0\over\tau_c} e^{-|t-t'|/\tau_c} ,   \label{gou}
\end{equation}
whether $t>t'$ or $t<t'$, where $\gamma_{0}$ is the friction and $\tau_{c}$ is the correlation time. With this friction memory kernel along with the approximation $F_e(t)={\mathcal K}QE_0\sin(\Omega t)$, we calculate the average value $\langle x(t)\rangle$, which is given by Eq. (\ref{mxb}) in App.~\ref{app:b} 1. Also, the work mean value is explicitly calculated in App.~\ref{app:b} 2, and given by Eq.~\ref{W2}. These expressions allow us to obtain the average of the total entropy production in the time interval $[0,\tau]$, as given by Eq.~\ref{mDst}, that is
\be \label{mDst-2}
	\langle \Delta s_{tot}\rangle={1\over T} \left(\langle W(\tau)\rangle - {1\over 2}k \langle x(\tau)\rangle^2 +\langle x(\tau) \rangle F_e(\tau)\right) .
\ee

Next, we explore the Markovian and non-Markovian time evolution of mean work and mean total entropy production as a function of the two parameters of the system: the oscillator frequency $\omega$, and the friction $\gamma_{0}$. For two values of the driving frequency, $\Omega=0.5,~1.5$, we analyze the overdamped and underdamped regimes. Notice that, in general, $\Omega$ controls how fast energy is injected with a driving time scale $\Omega^{-1}$, while the bath's memory determines how much of that energy can be stored before being dissipated at a rate of $\tau_c$, (which does not happen in the Markovian case, $\tau_c=0$). This time scales enter in competition with the characteristic time of the oscillator $\omega^{-1}$, determining the system response and the relevance of non-Markovian effects.\vspace{0.1cm}

The time behavior in the overdamped regime ($\gamma_{0}>\omega$) are shown in Fig.~\ref{fig:1}. First, panels (a)-(d) display the effect of weak driving $\Omega=0.5$. In the Markovian limit, panels (a) and (c), friction is able to compensate for driving, dissipating the energy injected. Therefore, both the average work and the entropy production exhibit an overall linear increase over time, with an oscillatory component superimposed. This behavior arises from the presence of an external force that continuously drives the system out of equilibrium. Because $\omega \gg \Omega$, the intrinsic dynamics of the oscillator are much faster than the variation of the external field, allowing the particle to adjust almost instantaneously to the force. Hence, the average position follows the field with a small phase lag, producing oscillations in both the work and the entropy production, reflecting the periodic injection of energy into the system and its partial release during each cycle. In contrast, in the non-Markovian case, panels (b) and (d), we observe an amplification of the observables in time. The bath’s memory introduces a delayed response that allows the system to temporarily store energy and partially reabsorb it, favoring energy accumulation, unlike instantaneous friction, which prevents it. Moreover, entropy production exhibits negative values at short times because the bath allows a transient backflow of energy into the system, a hallmark of non-Markovian dynamics. Additionally, the oscillations display an envelope resulting from the superposition of the external driving and the delayed response imposed by the memory kernel.

Increasing the driving field's frequency to $\Omega = 1.5$, leads us out of the quasi-static regime, as shown in Figs.~\ref{fig:1} (e)-(h). The the system is driven in a shorter time scale than the response induced by the bath's memory. Hence, the rapid change in the external field produces a more pronounced phase lag in the system response, and the oscillations become faster. Here, the Markovian limit still shows a limited response, since the injected energy is dissipated almost immediately. In contrast, in the non-Markovian regime, the growth of work is smaller both compared to the Markovian case and to the low-frequency regime. Since the external driving varies on a time scale shorter than the bath correlation time, the system response incorporates out-of-phase contributions. Then, the effective response loses correlation with the applied force, reducing the net energy transfer. 

The underdamped regime ($\omega > \gamma_0$) is shown in Fig.~\ref{fig:2}. Here, the dynamics are fully dominated by the system’s intrinsic oscillations, with an effective frequency $\sqrt{\omega^2 - (\gamma_0/2)^2}$, close to $\omega$ in the weak damping regime. Dissipation weakens, allowing the system to sustain large-amplitude oscillations and store energy in both kinetic and potential components. This explains why the oscillations in the work are much larger than in the overdamped regime, when the driving is small $\Omega=0.5$, as shown in Figs.~\ref{fig:2} (a)-(d). In the Markovian case, Figs.~\ref{fig:2} (a)-(c), even in the presence of an external force, the average work grows only slightly, indicating that the energy entering the system is not dissipated, but rather continuously exchanged within the system itself, with a small net accumulation of work. Consistently, the entropy production does not grow monotonically, but instead approximately oscillates around a constant value. We could say this is the case closest to reversibility, where work done is almost fully compensated by the change in internal energy. In the non-Markovian case, in Figs.~\ref{fig:2} (b)-(d), the entropy production can also become negative at short times, although less pronounced than in the overdamped regime, thanks to the dominance of the intrinsic oscillations over the backflow of energy enabled by memory. 

When $\Omega = 1.5$, the intrinsic frequency $\omega$ still governs the fast oscillations. Now, the rapidness of the external driving reduces the system's ability to follow it, as shown in Fig.~\ref{fig:2} (e)-(h). In the Markovian case, in Figs.~\ref{fig:2} (e) and (g), the response remains regular and oscillatory, with moderate growth in both average work and entropy production due to continuous but dissipative energy exchange with the environment. In contrast, in the non-Markovian case, in Figs.~\ref{fig:2} (f) and (h), delayed dissipation allows energy to accumulate more efficiently, due to the phase mismatch between the system and the external force. Consequently, the net energy transfer is reduced, leading to a smaller growth of both quantities, far weaker than in the overdamped regime. Additionally, the dynamics exhibits modulated oscillatory structures resembling wave packets, which develop after an initial transient regime, as the delayed contributions induced by the bath's memory become more pronounced in the system’s evolution. Therefore, although $\omega$ continues to determine the fast oscillations, the interplay between rapid driving and memory effects modifies the efficiency of energy transfer.

\section{Conclusions}
\label{sec:6}

Using the solution of the generalized Langevin equation along with Kubo's second fluctuation-dissipation theorem, we have been able to prove the validity of the detailed fluctuation theorem for total entropy production in the case of the Non-Markovian diffusive problem of an electron gas in a harmonic trap under the action of a time-dependent electric field. Here, it is assumed that the electric field influences not only the tagged Brownian particle but also the remaining bath particles. We have proposed that the induced force arising from this fact be added to the applied electric force, yielding a time-dependent effective force that drives the system out of equilibrium. If the detailed fluctuation theorem for the total entropy production is valid, then the validity of the integral fluctuation theorem holds immediately. We have studied and verified this result for two different, solvable time-dependent protocols. In the first case, we consider an arbitrary external time-dependent electric field; in the second case, the protocol includes a translation of the potential trap's minimum.

Then, we have taken the particular case of an oscillating electric field and an Ornstein-Uhlenbeck type friction memory kernel. If the oscillation frequency of each bath oscillator is much larger than the oscillation frequency of the applied electric field, then the effective electric force is proportional to the applied electric field. This condition supports our hypothesis that the effective force is responsible for driving the system out of equilibrium. 

Finally, we have explicitly calculated the average total entropy production and work as functions of the system's relevant parameters, discussing the dynamics in the overdamped and underdamped regimes, including the case of small external driving and when the system begins to leave the quasi-static regime.

\section*{Acknowledgements}
MABM and JIJA thank SNII-SECIHTI. KSRV thanks the ``SNII Research Assistant'' program from SECIHTI. MABM acknowledges financial support from CONAHCYT/SECIHTI No. CBF2023-2024-1765, the financial support from the Marcos Moshinsky Fellowship Program, and the PIPAIR 2024 project from the DAI-UAM.

\appendix

\section{Calculation of $\langle Wx\rangle-\langle W\rangle \langle x\rangle$ }
\label{app:a}

Here, we provide explicit calculations of the work-position correlation function necessary to obtain the variance of the total entropy production.
\ba
&&\langle WX\rangle=\langle Wx\rangle-\langle W\rangle \langle x\rangle\cr\cr
&=&-\left\langle\left(\int_0^{\tau} \dot F_e(t) \, x(t)\, dt\right) X(\tau)\right\rangle\cr\cr
&=&-\int_0^{\tau} \dot F_e(t) \langle[X(t) +\langle x\rangle] X(\tau)\rangle\, dt\cr\cr
&=&
-\int_0^{\tau} \dot F_e(t) \langle X(t)X(\tau)\rangle\, dt . \label{a1}
\ea
However, according to Eq.~\ref{corX3}  
\ba
\langle WX\rangle&=&-{k_BT\over k}
\int_0^{\tau}\dot F_e(t)  H(\tau-t)\, dt \cr\cr
&=&-{k_BT\over k}\bigg[ F_e(\tau)-k \int_0^{\tau} G(\tau-t) F_e(t) \,dt \bigg]\cr\cr
&=&-{k_BT\over k}[F_e(\tau)-k\langle x\rangle] . \label{a2} \ea

\section{Calculation of $\langle x(t)\rangle$, and $\langle W(\tau)\rangle$ for an Ornstein-Uhlenbeck type friction memory kernel and an oscillating electric field}
\label{app:b}

\subsection{Calculation of $\langle x(t)\rangle$}

According to Eq.~\ref{csolmx}, we need the $G(t)$ function for an OU-type friction memory kernel given by Eq. (\ref{gama2}). In this case, the Laplace transform of $G(t)$ becomes
\be
\hat G(s)={s+a \over s^3 + as^2 + bs + c }, \label{B1}
\ee
where $a=1/\tau_c$, ~$b=a\gamma_0+k$, ~$c=ak$. 
The denominator is a third-degree polynomial, which has three roots following the conditions that the discriminant $\Delta$ must satisfy, that is, $\Delta=0$, $\Delta<0$, and $\Delta>0$. In our work we choose the case $\Delta>0$, which allows us to obtain one real root $\lambda_1$ and two complex roots $\lambda_2$ and $\lambda_3$, such that 
\be
s^3 + as^2 + bs + c =(s-\lambda_1)(s-\lambda_2)(s-\lambda_3)=0 , \label{B2}
\ee
where 
\ba
\lambda_1&=&S_1+S_2-{a\over 3},\label{B3} \\
\lambda_2&=&-{(S_1+S_2)\over 2}-{a\over 3} + i{\sqrt{3\over 2}} (S_1-S_2) , \label{B4} \\
\lambda_3&=&- {(S_1+S_2)\over 2}-{a\over 3} - i{\sqrt{3}\over 2}(S_1-S_2), \label{B5}
\ea
and 
\ba
S_1&=&\sqrt[3]{R+\sqrt{Q^3+R^2}},\quad S_2=\sqrt[3]{R-\sqrt{Q^3+R^2}} \nonumber\\ \label{B6}\\
Q&=& {1\over 9} (3b-a^2)
\qquad R={1\over 54}(9ab -27c -2a^3) \label{B7}.
\ea
The discriminant is $\Delta=Q^3+R^2$, which satisfies 
\be
\Delta=-{a^2b^2\over108}+{b^3\over 27}+{c a^3\over 27}-{a b c\over 6}+{c^2\over 4} . \label{Del}
\ee
$G(t)$ is the inverse Laplace transform of $\hat G(s)$, i.e., 
\be
G(t)=
{\mathcal L}^{-1}\left\{{s+a\over (s-\lambda_1)(s-\lambda_2)(s-\lambda_3)}\right\}, \label{Gt1}
\ee
and after some algebra we get  
\ba
G(t)&=& {(a+\lambda_1) ~e^{\lambda_1 t}\over(\lambda_1-\lambda_2)(\lambda_1-\lambda_3)}- {(a+\lambda_2) ~e^{\lambda_2 t}\over(\lambda_1-\lambda_2)(\lambda_2-\lambda_3)}\cr\cr
&+& {(a+\lambda_3) ~e^{\lambda_3 t}\over(\lambda_1-\lambda_3)(\lambda_2-\lambda_3)} .  \label{Gt2}\ea
If we now define $p=-(S_1+S_2)+a/3$,
$q=(S_1+S_2)/2+a/3$, $\lambda=\sqrt{3}(S_1-S_2)/2$, thus $\lambda_1=-p$, $\lambda_2=-q+i \lambda$, $\lambda_3=-q-i\lambda$. Therefore
\be
G(t)=c_1\, e^{-pt} - {c_1 (d-q)\over\lambda} e^{-qt} \sin(\lambda t) 
-c_1 e^{-qt} \cos(\lambda t)  , \label{Ga}
\ee
where 
\ba
c_1&=&{a-p\over (q-p)^2+\lambda^2}={1\over 2q-p-d} , \cr\cr
 d&=&{a(2q-p)-q^2-\lambda^2\over a-p}.
\label{coef}\ea
Now we consider in particular an oscillating electric field. It has been shown in Sec.~\ref{sec:4} that $F_e(t)={\mathcal K}QE_0 \sin(\Omega t)$; so, the mean value of $x(t)$ according to Eqs.~\ref{csolmx} and~\ref{Ga} reads
\ba
{\langle x(t)\rangle\over R_0 c_1}&=& \int_0^t e^{-p(t-t')}   \sin(\Omega t') \,dt' \cr\cr
&-& {d-q\over\lambda}  \int_0^t e^{-q(t-t')} \sin[\lambda(t-t')]  \sin(\Omega t')\,dt' \cr\cr
&-& \int_0^t e^{-q(t-t')} \cos[\lambda(t-t')] \sin(\Omega t')\, dt' , \label{mxa} \ea
with $R_0={\mathcal K}QE_0 $. After a long algebra, we conclude that 
\ba
\langle x(t)\rangle&=&R_0 c_1\bigg\{\mathcal A\,e^{-pt} -[{\mathcal A}-{\mathbb C}] \cos(\Omega t)+ [{\mathcal B}+{\mathbb B}] \sin(\Omega t)\cr\cr
&-&{\mathbb C} e^{-qt}\, \cos(\lambda t)- {\mathbb D} e^{-qt}\,\sin(\lambda t)\bigg\} ,
 \label{mxb} \ea
where 
\ba
{\mathcal A}&=&{\Omega\over p^2+\Omega^2}, \qquad {\mathcal B}={p\over p^2+\Omega^2},\cr\cr
{\mathbb B}&=&{\mathcal E}+{\mathcal F} +{d-q\over\lambda}(\mathcal C+\mathcal D), \cr\cr
{\mathbb C}&=&{\mathcal C}-{\mathcal D} -{d-q\over\lambda}{(\mathcal E}-{\mathcal F)},\cr\cr
{\mathbb D}&=&{\mathcal E}-{\mathcal F} +{d-q\over\lambda} (\mathcal C}-{\mathcal D),\cr\cr 
{\mathcal C}&=&{\lambda+\Omega\over2[q^2+(\lambda+\Omega)^2]}, \qquad {\mathcal D}={ \lambda-\Omega\over 2[q^2+(\lambda-\Omega)^2]} ,\cr\cr
{\mathcal E}&=&{q\over 2[q^2+(\lambda+\Omega)^2]}, \qquad 
 {\mathcal F}={q\over 2[q^2+(\lambda- \Omega)^2]}. \nonumber  \\  \label{coffa}\ea

\subsection{Calculation of $\langle W(\tau)\rangle$}

Using Eqs.~\ref{mxb} and~\ref{mWt}, the work mean value can be written as  
\ba\label{W1}
 {\langle W\rangle\over R_0^2 \, c_1\Omega}&=&-{\mathcal A}\,\int_0^{\tau} e^{-pt}  \cos(\Omega t)\,dt \cr\cr
&+&[{\mathcal A}
-{\mathbb C}] \int_0^{\tau} \cos^2(\Omega t)\, dt \cr\cr
&-&[{\mathcal B}+{\mathbb B}]\int_0^{\tau} \sin(\Omega t) \cos(\Omega t)\, dt \cr\cr
&+&{\mathbb C} \int_0^{\tau} e^{-qt}\, \cos(\lambda t) \cos(\Omega t)\, dt \cr\cr
&+&{\mathbb D} \int_0^{\tau} e^{-qt}\,\sin(\lambda t) \cos(\Omega t)\, dt. 
\ea
After evaluating the integrals we arrive to 
\ba\label{W2}
&&{\langle W\rangle\over R_0^2 \, c_1\Omega}={1\over 2}[{\mathcal A}-{\mathbb C}]\, \tau+ {1\over 4\Omega}[{\mathcal A}-{\mathbb C}] \, \sin(2\Omega \tau) \cr\cr
&-&{{\mathcal A}\,p\over p^2+\Omega^2} [1- e^{-p\tau} \cos(\Omega \tau)]
-{{\mathcal A}\,\Omega\over p^2+\Omega^2}\,e^{-p\tau}\,\sin(\Omega \tau) \cr\cr
&-&{1\over 2\Omega}[{\mathcal B}+{\mathbb B}]
\, \sin^2(\Omega \tau)\cr\cr
&+&{ {\mathbb C}(\lambda+\Omega) -{\mathbb D}\,q\over 2[q^2+(\lambda+\Omega)^2]} \, e^{-q\tau} \sin[(\lambda+\Omega)\tau]\cr\cr
&+&{{\mathbb C}(\lambda-\Omega) -D\,q\over 2[q^2+(\lambda-\Omega)^2]}\, e^{-q\tau} \sin[(\lambda-\Omega)\tau]\cr\cr
&+&{{\mathbb C}\,q +{\mathbb D}(\lambda+\Omega)\over 2[q^2+(\lambda+\Omega)^2]}\bigg\{ 1- e^{-q\tau}\cos[(\lambda+\Omega)\tau]\bigg\} \cr\cr
&+&{{\mathbb C}\,q +{\mathbb D}(\lambda-\Omega)\over 2[q^2+(\lambda-\Omega)^2]}\bigg\{1-e^{-q\tau} \cos[(\lambda-\Omega)\tau]\bigg\} . \ea

\section{Markovian limit}

Here, we derive the Markovian limit of the response kernel $G(t)$ starting from its non-Markovian expression.

\subsection{Non-Markovian kernel}

The response kernel in the non-Markovian case can be written as
\begin{gather}
\label{eq:nucleo}
G(t)= c_1 e^{-p t}-c_1 \frac{d-q}{\lambda} e^{-q t}\sin(\lambda t)\\ \nonumber
-c_1 \frac{d-q}{\lambda} e^{-q t}\cos(\lambda t),
  \end{gather}
  where the parameters $p,q,\lambda,d$ and $c_1$ are determined by the poles of the Laplace transform of the kernel. The associated Laplace transform reads
\begin{equation}
\hat G(s)=\frac{s+a}{s^3+a s^2+b s+c}
=\\ \nonumber
\frac{s+a}{(s-\lambda_1)(s-\lambda_2)(s-\lambda_3)},
\end{equation}
whose poles are given by
\begin{equation}
\lambda_1=-p,\qquad
\lambda_2=-q+i\lambda,\qquad
\lambda_3=-q-i\lambda.
\end{equation}
Finally, for the characteristic polynomial
\begin{equation}
P(s)=s^3+a s^2+b s+c,
\end{equation}
the Viete relations yield
\begin{align} 
\lambda_1 + \lambda_2 + \lambda_3 &= -a \label{eq:viete1} \\ 
\lambda_1 \lambda_2 + \lambda_1 \lambda_3 + \lambda_2 \lambda_3 &= a \gamma_0 + \omega^2 \label{eq:viete2} \\ 
\lambda_1 \lambda_2 \lambda_3 &= -a \omega^2 \label{eq:viete3} 
\end{align}

\subsection{Scaling of the roots and the limit $a\to\infty$}

To study the Markovian limit $a\to\infty$, we analyze the scaling of the roots. Since the polynomial contains different powers of $a$, not all roots can scale in the same way. We therefore assume that one root scales linearly with $a$, i.e., $s=\alpha a$. By substituting this ansatz into $P(s)$ and retaining the leading terms in $a$, one finds that $\alpha^2(\alpha+1)=0$, which implies the existence of a root at $\alpha=-1$, so $s\sim -a$. Consequently,$\lambda_1=-p\sim -a$. This implies that the first term in the kernel Eq~\eqref{eq:nucleo} vanishes in the Markovian limit,
\begin{equation}
\lim_{a\to\infty} c_1 e^{-p t}=0.
\end{equation}
To better express the behavior of $p$, we write $p=a-\gamma_0+\mathcal{O}(a^{-1})$. Using the Viete relation in Eq.~\eqref{eq:viete1}, we obtain
\begin{equation}
-p-2q=-a \quad \Rightarrow \quad q=\frac{\gamma_0}{2}+\mathcal{O}(a^{-1}).
\end{equation}
Then, from the Viete relation \eqref{eq:viete3} and $\lambda_1=-p$, it follows that
\begin{equation}
\lambda_2\lambda_3=\frac{a\omega^2}{p}
=\omega^2+\mathcal{O}(a^{-1}).
\end{equation}
Since $\lambda_2\lambda_3=q^2+\lambda^2$, we finally obtain
\begin{equation}
\lambda=\sqrt{\omega^2-\frac{\gamma_0^2}{4}}.
\end{equation}

\subsection{Limit of the kernel}

The asymptotic behavior of the remaining coefficients is
\begin{align}
d &= -\frac{a^2}{\gamma_0}+2a-\frac{\omega^2}{\gamma_0}+\mathcal{O}(a^{-1}),\,\,\,\,\, c_1\sim \frac{\gamma_0}{a^2},
\end{align}
which implies
\begin{equation}
\lim_{a\to\infty} c_1(d-q)=-1.
\end{equation}
Substituting these results into Eq.~\eqref{eq:nucleo} and taking the limit $a\to\infty$, we obtain the Markovian kernel
\begin{equation}
G(t)=\frac{1}{\lambda} e^{-\frac{\gamma_0 t}{2}}\sin(\lambda t),
\end{equation}
valid in the underdamped regime $2\omega>\gamma_0$.
In the overdamped regime, defining
\begin{equation}
\lambda=\frac{i\beta}{2}, \qquad \beta=\sqrt{\gamma_0^2-4\omega^2}.
\end{equation}
The kernel takes the equivalent form
\begin{equation}
G(t)=\frac{2}{\beta} e^{-\frac{\gamma_0 t}{2}}\sinh\left(\frac{\beta t}{2}\right).
\end{equation}
This result coincides with the response kernel obtained from a Markovian description of the system.

\bibliography{biblio.bib}

@misc{Argun2016,
  author  = {A. Argun and A. R. Moradi and E. Pince and G. B. Bagci and G. Volpe},
  title   = {Experimental evidence of the failure of Jarzynski equality in active baths},
  howpublished = {arXiv:1601.01123},
  year    = {2016},
  doi     = {10.48550/arXiv.1601.01123}
}

@article{Bochkov1981,
title = {Nonlinear fluctuation-dissipation relations and stochastic models in nonequilibrium thermodynamics: I. Generalized fluctuation-dissipation theorem},
journal = {Physica A: Statistical Mechanics and its Applications},
volume = {106},
number = {3},
pages = {443-479},
year = {1981},
issn = {0378-4371},
doi = {https://doi.org/10.1016/0378-4371(81)90122-9},
url = {https://www.sciencedirect.com/science/article/pii/0378437181901229},
author = {G.N. Bochkov and Yu.E. Kuzovlev}
}

@book{Binder2018,
  author    = {F. Binder and L. A. Correa and C. Gogolin and J. Anders and G. Adesso},
  title     = {Thermodynamics in the quantum regime},
  booktitle = {Fundamental Theories of Physics},
  volume    = {195},
   publisher = {Springer Cham},
  address   = {Switzerland},
  year      = {2018}
}

@article{Crooks1999,
  author  = {G. E. Crooks},
  title   = {Entropy production fluctuation theorem and the nonequilibrium work relation for free energy differences},
  journal = {Phys. Rev. E},
  volume  = {60},
  pages   = {2721},
  year    = {1999},
  doi = {https://doi.org/10.1103/PhysRevE.60.2721} 
}

@article{Chaki2018,
  author  = {Subhasish Chaki and Rajarshi Chakrabarti},
  title   = {Entropy production and work fluctuation relations for a single particle in active bath},
  journal = {Physica A},
  volume  = {511},
  pages   = {302},
  year    = {2018},
  doi = {10.1016/j.physa.2018.07.055}
}

@article{Cui2018,
  author  = {Bingyu Cui and Alessio Zaccone},
  title   = {Generalized Langevin equation and fluctuation-dissipation theorem for particle-bath systems in external oscillating fields},
  journal = {Phys. Rev. E},
  volume  = {97},
  pages   = {060102},
  year    = {2018},
  doi     = {10.1103/PhysRevE.97.060102}
}

@article{Evans1993,
  title = {Probability of second law violations in shearing steady states},
  author = {Evans, Denis J. and Cohen, E. G. D. and Morriss, G. P.},
  journal = {Phys. Rev. Lett.},
  volume = {71},
  issue = {15},
  pages = {2401--2404},
  numpages = {0},
  year = {1993},
  month = {Oct},
  publisher = {American Physical Society},
  doi = {10.1103/PhysRevLett.71.2401},
  url = {https://link.aps.org/doi/10.1103/PhysRevLett.71.2401}
}

@incollection{Funo2018,
  author      = "K. Funo and M. Ueda and T. Sagawa",
  title       = "Quantum fluctuation theorems",
  editor      = "F. Binder and L. A. Correa and C. Gogolin and J. Anders and G. Adesso",
  booktitle   = "Thermodynamics in the Quantum Regime",
  publisher   = "Springer Cham",
  address     = "Switzerland",
  year        = "2018",
  pages       = "249-274",
  chapter     = 10,
}

@article{Gallavotti1995,
  author  = {G. Gallavotti and E. G. D. Cohen},
  title   = {Dynamical ensembles in nonequilibrium statistical mechanics},
  journal = {Phys. Rev. Lett.},
  volume  = {74},
  pages   = {2694},
  year    = {1995},
  doi  ={10.1103/PhysRevLett.74.2694}
}

@article{Goswami2019,
  author  = {Koushik Goswami},
  title   = {Work fluctuation relations for a dragged Brownian particle in active bath},
  journal = {Physica A},
  volume  = {525},
  pages   = {223},
  year    = {2019},
  doi = {10.1016/j.physa.2019.03.050}
}

@article{Ghosh2017,
  author  = {B. Ghosh and S. Chaudhury},
  title   = {Fluctuation theorems for total entropy production in generalized Langevin systems},
  journal = {Physica A},
  volume  = {466},
  pages   = {133},
  year    = {2017},
  doi = {https://doi.org/10.1016/j.physa.2016.09.001}
}

@article{Hack2023,
  author  = {P. Hack and C. Lindig-Leon and S. Gottwald and D. A. Braun},
  title   = {Thermodynamic fluctuation theorems govern human sensorimotor learning},
  journal = {Sci. Rep.},
  volume  = {13},
  pages   = {869},
  year    = {2023},
  doi     = {10.1038/s41598-023-27736-8}
}

@Article{Hayashi2018,
author={Hayashi, Kumiko},
title={Application of the fluctuation theorem to motor proteins: from F1-ATPase to axonal cargo transport by kinesin and dynein},
journal={Biophysical Reviews},
year={2018},
month={Oct},
day={01},
volume={10},
number={5},
pages={1311-1321},
issn={1867-2469},
doi={10.1007/s12551-018-0440-5},
url={https://doi.org/10.1007/s12551-018-0440-5}
}

@article{Jarzynski1997,
  author  = {C. Jarzynski},
  title   = {Equilibrium free-energy differences from nonequilibrium measurements: A master-equation approach},
  journal = {Phys. Rev. E},
  volume  = {56},
  pages   = {5018},
  year    = {1997},
  doi= {10.1103/PhysRevE.56.5018}
}

@article{Jimenez2013,
  author  = {J. I. Jim{\'e}nez-Aquino and R. M. Velasco},
  title   = {Power fluctuation theorem for a Brownian harmonic oscillator},
  journal = {Phys. Rev. E},
  volume  = {87},
  pages   = {022112},
  year    = {2013},
  doi     = {10.1103/PhysRevE.87.022112}
}

@article{Jimenez2010,
  author  = {J. I. Jim{\'e}nez-Aquino},
  title   = {Entropy production theorem for a charged particle in an electromagnetic field},
  journal = {Phys. Rev. E},
  volume  = {82},
  pages   = {051118},
  year    = {2010},
  doi = {https://doi.org/10.1103/PhysRevE.82.051118}
}

@article{Jimenez2015,
  author  = {J. I. Jim{\'e}nez-Aquino},
  title   = {Non-Markovian work fluctuation theorem in crossed electric and magnetic fields},
  journal = {Phys. Rev. E},
  volume  = {92},
  pages   = {022149},
  year    = {2015},
  doi = {https://doi.org/10.1103/PhysRevE.92.022149}
}

@article{Kurchan1999,
  author  = {J. Kurchan},
  title   = {Fluctuation theorem for stochastic dynamics},
  journal = {J. Phys. A: Math. Gen.},
  volume  = {31},
  pages   = {3719},
  year    = {1999},
  doi = {10.1088/0305-4470/31/16/003}
}

@article{Lisy2019,
title = {Generalized Langevin equation and the fluctuation-dissipation theorem for particle-bath systems in a harmonic field},
journal = {Results in Physics},
volume = {12},
pages = {1212-1213},
year = {2019},
issn = {2211-3797},
doi = {https://doi.org/10.1016/j.rinp.2019.01.003},
url = {https://www.sciencedirect.com/science/article/pii/S2211379718328195},
author = {Vladimír Lisý and Jana Tóthová}}

@article{Mahault2022,
  author  = {Benoit Mahault and Evelyn Tang and Ramin Golestanian},
  title   = {A topological fluctuation theorem},
  journal = {Nat. Commun.},
  volume  = {13},
  pages   = {3036},
  year    = {2022},
  doi     = {10.1038/s41467-022-30644-6}
}

@article{Mai2007,
  author  = {T. Mai and A. Dhar},
  title   = {Nonequilibrium work fluctuations for oscillators in non-Markovian baths},
  journal = {Phys. Rev. E},
  volume  = {75},
  pages   = {061101},
  year    = {2007},
  doi = {https://doi.org/10.1103/PhysRevE.75.061101}
}

@article{Noh2012,
  author  = {Jae Dong Noh and Jong-Min Park},
  title   = {Fluctuation Relation for Heat},
  journal = {Phys. Rev. Lett.},
  volume  = {108},
  pages   = {240603},
  year    = {2012},
  doi     = {10.1103/PhysRevLett.108.240603}
}

@article{Narinder2021,
  author  = {N. Narinder and P. Shuvojit and C. Bechinger},
  title   = {Work fluctuation relation of an active Brownian particle in a viscoelastic fluid},
  journal = {Phys. Rev. E},
  volume  = {104},
  pages   = {034605},
  year    = {2021},
  doi = {https://doi.org/10.1103/PhysRevE.104.034605}
}

@article{Ohkuma2007,
  author  = {T. Ohkuma and T. Ohta},
  title   = {Fluctuation theorems for non-linear generalized Langevin systems},
  journal = {J. Stat. Mech.},
  pages   = {P10010},
  year    = {2007},
  doi = {10.1088/1742-5468/2007/10/P10010}
}

@article{Park2023,
  author  = {Jung Jun Park and Hyunchul Nha},
  title   = {Fluctuation Theorem for Information Thermodynamics of Quantum Correlated Systems},
  journal = {Entropy},
  volume  = {25},
  pages   = {165},
  year    = {2023},
  doi     = {10.3390/e25010165}
}

@article{Rao2018,
  author  = {Ricardo Rao and Massimiliano Esposito},
  title   = {Detailed Fluctuation Theorems: A Unifying Perspective},
  journal = {Entropy},
  volume  = {20},
  pages   = {635},
  year    = {2018},
  doi     = {10.3390/e20090635}
}

@article{Semeraro2024,
  author  = {Massimiliano Semeraro and Antonio Suma and Giuseppe Negro},
  title   = {Fluctuation Theorems for Heat Exchanges between Passive and Active Baths},
  journal = {Entropy},
  volume  = {26},
  pages   = {439},
  year    = {2024},
  doi     = {10.3390/e2606439}
}

@article{Sidajaya2025,
  author  = {Peter Sidajaya and Jovan Hsuen Khai Low and Clive Cenxin Aw and Valerio Scarani},
  title   = {Emergence of fluctuation relations in UNO},
  journal = {Phys. Rev. Res.},
  volume  = {7},
  pages   = {023054},
  year    = {2025},
  doi     = {10.1103/PhysRevResearch.7.023054}
}

@article{Seifert2005,
  title = {Entropy Production along a Stochastic Trajectory and an Integral Fluctuation Theorem},
  author = {Seifert, Udo},
  journal = {Phys. Rev. Lett.},
  volume = {95},
  issue = {4},
  pages = {040602},
  numpages = {4},
  year = {2005},
  month = {Jul},
  publisher = {American Physical Society},
  doi = {10.1103/PhysRevLett.95.040602},
  url = {https://link.aps.org/doi/10.1103/PhysRevLett.95.040602}
}

@article{Saha2009,
  author  = {A. Saha and S. Lahiri and A. M. Jayannavar},
  title   = {Entropy production and fluctuation theorems in driven mesoscopic systems},
  journal = {Phys. Rev. E},
  volume  = {77},
  pages   = {011117},
  year    = {2009},
  doi = {https://doi.org/10.1103/PhysRevE.80.011117}
}

@book{Sekimoto2010,
  author    = {K. Sekimoto},
  title     = {Stochastic Energetics},
  series    = {Lecture Notes in Physics},
  volume    = {799},
  publisher = {Springer},
  address   = {Heidelberg, Germany},
  year      = {2010}
}

@article{vanZon2004,
  author  = {R. van Zon and S. Ciliberto and E. G. D. Cohen},
  title   = {Power and heat fluctuation theorem for electric circuit},
  journal = {Phys. Rev. Lett.},
  volume  = {92},
  pages   = {130601},
  year    = {2004},
  doi     = {10.1103/PhysRevLett.92.130601}
}

@article{Zeng2021,
  author  = {Qian Zeng and Jin Wang},
  title   = {New fluctuation theorems on Maxwell's demon},
  journal = {Sci. Adv.},
  volume  = {7},
  pages   = {e1807},
  year    = {2021},
  doi = {10.1126/sciadv.abf1807}
}

@book{Zwanzig2001,
  author    = {R. Zwanzig},
  title     = {Nonequilibrium Statistical Mechanics},
  publisher = {Oxford University Press},
  address   = {New York},
  year      = {2001}
}

\end{document}